\documentclass[epj]{svjour}
\usepackage{latexsym}
\usepackage{graphicx}
\begin{document}
\title{The Bethe-Salpeter Equation and the Low Energy Theorems for 
$\pi N$ Scattering}
\author{I.R.~Afnan\inst{1} \and A.D.~Lahiff\inst{2}
\thanks{The authors would like to thank the Australian Research 
Council for their financial support during the course of this work. 
The work of A.D.L. was supported in part by a grant from the National 
Science and Engineering Research Council of Canada.}%
}                     

\institute{School of Chemistry, Physics \& Earth Sciences, 
Flinders University, GPO Box 2100, Adelaide 
5001, Australia \and TRIUMF, 4004 Wesbrook Mall,Vancouver, BC, Canada 
V6T~2A3 }
\date{Received: 30/9/2002}
%
\abstract{
The Bethe-Salpeter (BS) amplitude for $\pi N$ scattering is evaluated at 
the off mass shell points corresponding to the Low Energy Theorems (LET) 
based on PCAC and current algebra. The results suggest a way of maintaining
constructing between BS equation and LET.
}
     
\maketitle
\section{Introduction}
\label{intro}
The formulations of  $\pi N$ scattering can be divided into two 
approaches. On the one hand, we have the Effective Field Theory (EFT) 
approach where the emphasis is on preserving the symmetries of QCD. 
This is achieved by the expansion for the amplitude in powers of the 
pion mass or external momenta divided by a typical QCD cut-off of 
$\approx 1$~GeV. For the $\pi N$ system the commonly used EFT is 
Chiral Perturbation Theory (ChPT) \cite{GL84} which results in an 
off-mass-shell amplitude that is consistent with the Low Energy 
Theorems (LET). These LET are based on current algebra and PCAC. On 
the other hand, we have the more traditional approach of using the 
two-body scattering equation (\textit{e g.} the Bethe-Salpeter 
equation) in which the potential is based on $s$-, $t$-, and 
$u$-channel pole diagrams derived from a chirally invariant 
Lagrangian. In this case unitarity is the main uniting feature which 
allows the examination of $\pi N$ scattering at higher energies. 

\section{The Bethe-Salpeter Amplitude}\label{sec:1}

\begin{figure*}[t]
\centering\includegraphics[width=16cm,angle=-90]{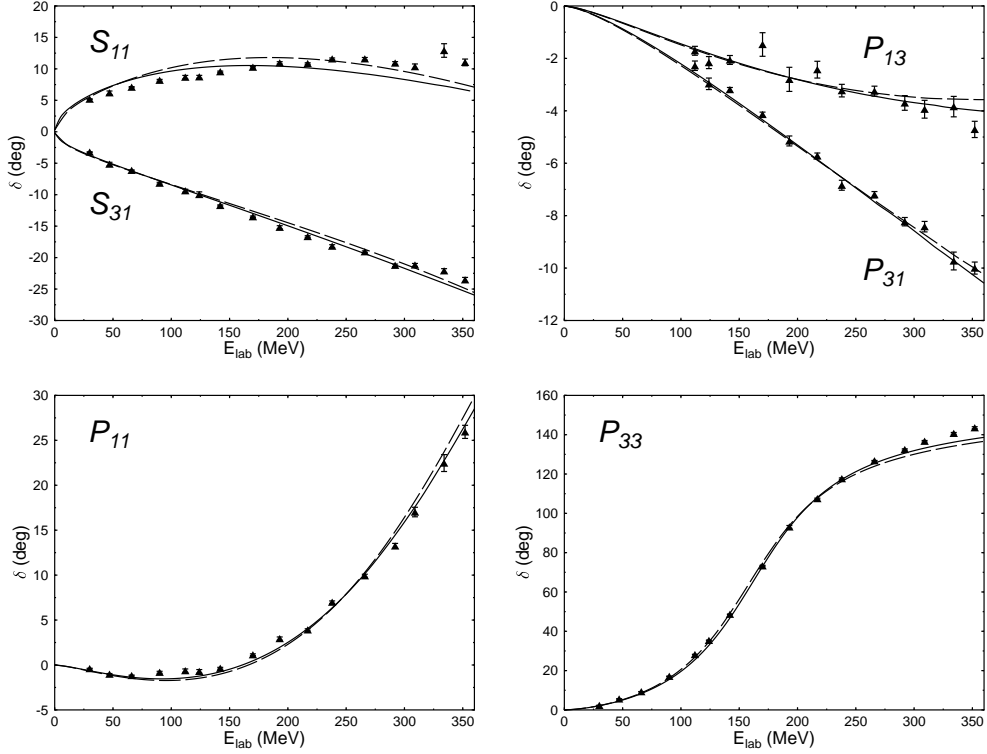}
\vspace{-5.0cm}
    \caption{The $S$ and $P$ wave phase shifts for the form factors
type I (solid line) and II (dashed line), the data is SM95\cite{A95}}\label{fig1}
\end{figure*}

With the advent of solutions to the Bethe-Salpeter (BS) equation for 
the off-mass-shell $\pi N$ amplitude,\cite{LA99} it is now possible 
to compare the results of the traditional approach based on two-body 
scattering, with the LET. Here we will present solutions to the BS 
equation based on a potential derived from the chirally invariant 
Lagrangian.\cite{LA99} 
\begin{eqnarray}
{\cal L}_{\rm int} &=&\frac{g_{\pi NN}}{2m_N}\bar{\psi}_N\gamma_5\gamma^\mu
\vec{\tau}\cdot\partial_\mu\vec{\pi}\psi_N
\nonumber\\
&&+\frac{f_{\pi N\Delta}}{m_\pi}\bar{\psi}^\mu_\Delta(g_{\mu\nu} + x_\Delta
\gamma_\mu\gamma_\nu)\vec{T}\psi_N\cdot\partial^\nu\vec{\pi} + h.c.
\nonumber\\
&&+g_{\rho NN}\bar{\psi}_N\frac{1}{2}\vec{\tau}
\cdot\left(\gamma^\mu\vec{\rho}_\mu
+\frac{\kappa_\rho}{2m_N}\sigma^{\mu\nu}\partial_\mu\vec{\rho}_\nu\right)
\psi_N\nonumber\\
&&+g_{\rho\pi\pi}\vec{\rho}_\mu\cdot(\partial^\mu\vec{\pi}\times\vec{\pi})
\nonumber\\
&&+g_{\sigma NN}\bar{\psi}_N\psi_N\sigma 
\nonumber\\
&&+ \frac{g_{\sigma\pi\pi}}{2m_\pi}
\sigma\partial_\mu\vec{\pi}\cdot\partial^\mu\vec{\pi}
\label{eq:1}
\end{eqnarray}
The tree level diagrams that contribute to the potential 
include the $s$- and $u$-channel diagrams with nucleon and $\Delta$ 
poles, and $t$-channel  diagrams with $\rho$ and $\sigma$ poles.  To 
solve the four-dimensional BS equation, we have introduced form 
factors for each of the vertices in the Lagrangian. Here we 
considered two classes of form factors
\begin{equation}
f_{\alpha\beta\gamma}(p_\alpha^2,p_\beta^2,p_\gamma^2) = 
f_\alpha(p_\alpha^2)\,
f_\beta(p_\beta^2)\,f_\gamma(p_\gamma^2)\quad\mbox{(Type I)}          
\label{eq:2}
\end{equation}
and
\begin{equation}
f_{\alpha\beta\gamma}(p_\alpha^2,p_\beta^2,p_\gamma^2) = 
f_\pi(p_\pi^2)
\qquad \qquad\qquad\mbox{(Type 
II)}                                                                              
\label{eq:3}
\end{equation}
where
\begin{equation}
f_\alpha(p_\alpha^2) = \left(\frac{\Lambda_\alpha^2 - m_\alpha^2}
{\Lambda_\alpha^2-p_\alpha^2}\right)^{n_\alpha}      \ .  
\qquad                    \label{eq:4}
\end{equation}
In the above, $m_\alpha$ and $\Lambda_\alpha$ are the mass and 
cut off associated with the hadron $\alpha$. The parameters in the 
Lagrangian are adjusted to fit the $s$- and $p$-wave scattering data 
up to pion laboratory energy of 300~MeV. In Fig.~\ref{fig1} we 
present a fit to the data for the form factors type I and II(n=4).
Similar fits are achieved for the form factors II(n=2) and II(n=10).
Here, the form factors determine the off mass shell behaviour of the BS 
amplitudes. In this way we can vary the off mass shell amplitudes 
when comparing the results of the BS equations with those based 
on current algebra and PCAC.

\section{The Low Energy Theorems}\label{sec:2}

The LET theorems for $\pi N$ scattering are based on current algebra 
and PCAC. The latter is implemented by defining the pion field, 
$\pi^a$, in terms of the derivative of the axial vector current, 
\textit{i e.} $\partial^\mu\,A_\mu^a = f_\pi\,m_\pi^2 \ \pi^a$.  
This allows us to write the $\pi N$ amplitude in terms of the commutation 
relation of the currents. In this way we can use current algebra to
determine the $\pi N$ amplitude in the soft pion limit. 

The $\pi N$ amplitude with off shell pion can be written as
\begin{equation}
T^{ab}_{\pi N} = \bar{u}(p')\left\{T^{(+)}\delta_{ab} + 
{\scriptstyle \frac{1}{2}}
\left[\tau_a,\tau_b\right]T^{(-)}\right\}u(p)\ ,\label{eq:5}
\end{equation}
where $p$ ($p'$) is the initial (final) on-shell nucleon momentum, and
\begin{equation}
T^{(\pm)}=D^{(\pm)} + \frac{i}{2m}\sigma^{\mu\nu}q_\mu q^{\prime}_\nu 
B^{(\pm)}\ .                     \label{eq:6}
\end{equation}
Here the $(+)$, $(-)$ refer to isospin even and odd components of the
amplitude, and $q$ ($q'$) are the off mass shell pion initial (final)
momentum. The amplitudes $D$ and $B$ with the pion off mass shell are a
function of $\nu$, $\nu_B$, $q^2$ and $q'^2$, \textit{i.e.}
$D^{(\pm)}= D^{(\pm)}(\nu,\nu_B,q^2,q'^2)$ with $\nu=\frac{1}{4m}(s-u)$
and $\nu_B$ being the value of $\nu$ at the $s$ channel nucleon pole,
and $s$ and $u$ being the standard Mandelstam variables. Current
algebra and PCAC can impose constraints on these off mass shell 
amplitudes. In particular, we can write the isospin even amplitude 
with the nucleon pole subtracted \textit{i e.} $\tilde{D}^{\,+}
(\nu,\nu_B,q^2,q'^2)$ at three off shell points. The amplitude at 
Weinberg~(W)\cite{W66}, Adler~(A)\cite{A65} and the 
Cheng-Dashen~(CD)\cite{CD71} points are:
\begin{eqnarray}
\tilde{D}^+(0,0,0,0) &=& 
-\frac{\sigma_{\pi N}(0)}{f_\pi^2}\ \label{eq:7}\\
\tilde{D}^+(0,0,m_\pi^2,0) &=& 0  = 
\tilde{D}^+(0,0,0,m_\pi^2) \  \label{eq:8}\\
\tilde{D}^+(0,0,m_\pi^2,m_\pi^2) &=&
\frac{\sigma_{\pi N}(0)}{f_\pi^2} + {\cal O}(m_\pi^4)
+ \cdots \ ,\label{eq:9}
\end{eqnarray}
respectively. Here we observe that the amplitude at the Weinberg and 
Cheng-Dashen points are opposite  in sign, while that at the Adler 
point is zero. Since the sigma term $\sigma_{\pi N}(0)$ is a measure of
chiral symmetry breaking \textit{i.e.} 
\begin{equation}
\sigma_{\pi N}(0) = \frac{1}{2}\sum_{a=1}^3\langle N(p)|\,
[Q^5_a,[Q^5_a,H]]|N(p)\rangle\ ,              \label{eq:10}
\end{equation}
in the absence of any mechanism for chiral symmetry breaking, the 
$\pi N$ amplitude is zero at all three points.

\section{Results}\label{sec:3}

To examine the variation in the off mass shell BS amplitude when
comparing with the results from the LET, we have considered four
possible form factors for our potential. In Table~\ref{table1} 
we have the parameters
that give the optimum fit to the data up to pion energy of 300~MeV for
the form factor types I(n=1), II(n=2), and II(n=4) and II(n=10). Also
included in the table are the equivalent cut off mass for a monopole
$\Lambda^R_\pi$, and the difference in the form factor at the pion 
pole and at $q^2=0$, \textit{i.e.} $\Delta_\pi=1-f^R_\pi(0)$. Here, $R$
refers to the fact that these quantities are calculated for the
renormalised form factor. From the table we observe that the dressing of the
nucleon and $\Delta$ is substantially more for type I form factors than
is the case for type II form factors. At the same time the type I form
factors give a value for $\Delta_\pi$ that is closer to the commonly
accepted value of $3\%$ from the Goldberger-Treiman relation.
\begin{table}[h]
\caption{The coupling constants and masses for the optimum fit to the
data for different choices for the form factors. All coupling
constants are $g^2/4\pi$.}\label{table1}
\begin{tabular}{lcccc}
\hline\noalign{\smallskip} 
               & I(n=1) & II(n=2) & II(n=4) & II(n=10)\\
\noalign{\smallskip}\hline\noalign{\smallskip}
$g_{\pi NN}^{(0)2}$ & 1.80 & 4.23  & 4.68 & 5.98 \\
$f^{(0)2}_{\pi n\Delta}$ & 0.37 & 0.17 & 0.20 & 0.196 \\
$x_\Delta$ & -0.11  & -0.13  & -0.24 & -0.18 \\
$g_{\rho NN}g_{\rho\pi\pi}$ & 2.88 & 2.67 & 2.63 & 2.80 \\
$\kappa_\rho$ & 2.66  & 2.18 & 2.03 & 2.15 \\
$g_{\sigma\pi\pi}g_{\sigma NN}$ & -0.41 & 0.86 & 0.39 & 0.48 \\
$m^{(0)}_N$ & 1.34 & 1.18 & 1.14 & 1.11 \\
$m^{(0)}_\Delta$ & 2.305 & 1.495 & 1.492 & 1.435 \\
$m_\sigma$ & 0.65 & 0.88 & 0.62 & 0.64 \\
\noalign{\smallskip}\hline\noalign{\smallskip}
$\Lambda_\pi^R$ & 1.22 & 0.874 & 0.868 & 0.822 \\
$\Delta_\pi$ & 1.3\% & 2.47\% & 2.51\% & 2.79\% \\
\noalign{\smallskip}\hline
\end{tabular}
\end{table}

In Table~\ref{table2} we present the off shell amplitude resulting from the 
solution of the BS equations at the Adler (A), Weinberg (W), and the 
Cheng-Dashen(CD) points for different choices for our form factor. Also 
included are the values for the $\pi N$ $\sigma$-term $\sigma_{\pi 
N}(0)$ and the isospin even scattering length $a^+$. Here we observe 
that:
\begin{table}[h]
    \caption{The BS amplitude at the Adler\cite{A65}, Weinberg\cite{W66} 
    and Cheng-Dashen\cite{CD71} points in units of $m_{\pi}^{-1}$ for 
    different form factors. Also included are the $\sigma$-term 
    $\sigma_{\pi N}(0)$ and the isospin even $S$-wave scattering 
    length.}\label{table2}
    \begin{tabular}{lccccc}
	\hline\noalign{\smallskip}
	Model & A & W & CD & $\sigma_{\pi N}(0)$ & $a^{+}$ \\
	\noalign{\smallskip}\hline\noalign{\smallskip}
	I        & 0.366 & 0.355 & 0.379 & 23.8 & -0.025 \\
	II $n=2$ & 0.0949      & 0.102     & 0.106   &  6.64 & -0.05 \\
        II $n=4$ & 0.0411      & 0.0462    & 0.0494  & 3.10  & -0.048\\
        II $n=10$& 0.0180      & 0.0218    & 0.0259  & 1.62  & -0.049\\
	\noalign{\smallskip}\hline
    \end{tabular}
\end{table}
\begin{enumerate}
    \item The off mass shell amplitude is sensitive to the choice of 
    cut off form factor, and in particular, model I gives a larger 
    $\sigma$-term than model II.
    \item The amplitude at the three off mass shell points are 
    approximately the same. This is in contrast to the fact that the 
    amplitude at the Weinberg and Cheng-Dashen points are equal and 
    opposite in sign.
    \item Finally, the amplitude at the Adler point is not zero.
\end{enumerate}
To understand the difference between the amplitude resulting from the 
solution of the BS equation and the LET, we first examined the Born 
amplitude for the $\pi N$ potentials considered. Here we found that the 
Born amplitude at all three points is zero, consistent with the 
requirement of chiral symmetry conservation. (The $\sigma$-term is a 
measure of chiral symmetry breaking). This suggests that the higher 
terms in the multiple scattering series give all the contribution to 
the amplitude at the three off mass shell points. To examine the 
source of this chiral symmetry breaking contribution, we have 
examined the contribution of each term in the potential to the
amplitude at the Adler, Weinberg and Cheng-Dashen points. This 
exercise revealed that the source of chiral symmetry breaking is 
the higher order multiple scattering $t$-channel $\rho$ exchange. In 
particular, it is the $\rho\pi\pi$ term in the Lagrangian that gives 
rise to the symmetry breaking.  Since this $\rho\pi\pi$ Lagrangian is 
of the form $g_{\rho\pi\pi}\,\vec{\rho}_{\mu}\cdot(\partial^{\mu}\vec{\pi}
\times\vec{\pi})$, then the only time this term in the Lagrangian 
contributes to chiral symmetry breaking is when the external pion is 
represented by $\vec{\pi}$ rather than $\partial^{\mu}\vec{\pi}$. It 
is possible to restore chiral symmetry to the amplitude by introducing 
a symmetric form for the coupling of the $\rho$ to the $\pi$, 
\textit{e.g.} $(\partial_{\mu}\rho_{\nu}-\partial_{\nu}\rho_{\mu})
\cdot(\partial^{\mu}\vec{\pi}\times\partial^{\nu}\vec{\pi})$\cite{WZ67}, 
which is equivalent on the mass shell to the form employed in the present 
calculation. The resultant amplitude would then satisfy chiral
symmetry. Chiral symmetry breaking could then be introduced in a 
controlled form via a non-derivative $\sigma\pi\pi$ coupling.

\section{Conclusion}\label{sec:4}

In the above analysis we have established that the Bethe-Salpeter 
amplitude for a Lagrangian that  satisfies chiral symmetry is 
inconsistent with the Low Energy Theorems of current algebra. The 
source of the disagreement is the mode of chiral symmetry breaking via 
the higher order multiple scattering of $t$-channel $\rho$ exchange. 
This could be overcome with the introduction of a symmetric $\rho\pi\pi$ 
Lagrangian, and a mode of chiral symmetry breaking that is consistent with 
the LET.

\end{document}